\documentclass[a4paper,11pt]{article}
\usepackage{jinstpub} 
\usepackage{lineno}

\usepackage[version=3]{mhchem} 
\usepackage{siunitx}
\usepackage{natbib}

\title{Commissioning and first results from the Cold Radon Emanation Facility}

\author[a,b,1]{E. Perry,\note{Corresponding author.}}
\author[a]{C. Ghag,}
\author[c]{M.G.D.van der Grinten,}
\author[a]{R. Saakyan,}
\author[a]{A. Stevens,}
\author[c]{M. Tucker,}
\author[a]{U. Utku,}
\affiliation[a]{University College London (UCL), Department of Physics and Astronomy, London WC1E 6BT, UK}
\affiliation[b]{Lawrence Berkeley National Laboratory, Berkeley, CA 94720-8099}
\affiliation[c]{STFC Rutherford Appleton Laboratory (RAL), Didcot, OX11 0QX, UK}

\emailAdd{eaperry@lbl.gov}

\abstract{Radon emanation from detector materials is a critical background for next-generation rare event searches, in particular those using noble liquid targets. While highly sensitive screening facilities mitigate this risk prior to detector construction, room-temperature assays often fail to predict cold emanation rates due to temperature-dependent diffusion suppression. The Cold Radon Emanation Facility at Rutherford Appleton Laboratory addresses this by performing high-sensitivity assays at detector operating temperatures. It includes a \SI{2.7}{L} small-sample chamber, a \SI{200}{L} chamber for large as-built components operated with a radon concentration line, a cryogenic infrastructure enabling measurements of emanation as a function of temperature, and an electrostatic radon detector, which achieves a minimum detectable activity of $\sim$\SI{0.05}{mBq} at 90\%~CL. Commissioning results and initial comparative assays are reported, including a preliminary indication of a factor of ~2 suppression of $^{222}$Rn emanation in a titanium sample at cryogenic temperatures. This result, obtained as part of commissioning measurements, illustrates the potential impact of temperature-dependent effects and underscores the importance of in-situ cold assays for future noble liquid detector components.}


\begin{document}
\maketitle
\flushbottom

\section{Introduction}
\label{sec:Introduction}


Despite major advances in material screening and cleanliness, radon emanation remains a significant background in current and next-generation rare-event experiments. The isotopes of primary concern are $^{222}$Rn (half-life 3.82 days) and $^{220}$Rn (half-life 55.8 s), originating from the $^{238}$U and $^{232}$Th decay chains, respectively. The impact of $^{220}$Rn is typically reduced due to its relatively short half-life, which limits its transport into active detector volumes. Radon emanation from materials arises through two primary mechanisms: recoil, in which radon atoms produced in radium decay are ejected from surfaces with keV-scale energies, and diffusion, whereby radon migrates through the material prior to decay. Diffusion depends on material properties such as temperature, density, surface conditions and lattice structure.

Over the past decade, substantial progress has been made in reducing radon levels through improved screening campaigns and in-line removal systems~\cite{ScreeningPaper2020, Murra_2022}. Nevertheless, residual radon continues to limit the sensitivity of current experiments ($<$~\SI{5}{\mu Bq/kg}~\cite{Aalbers_2023, Aprile_2023}). Future detectors, such as XLZD, will require a further order-of-magnitude reduction to achieve sub-dominant radon backgrounds and fully exploit their discovery potential~\cite{Aalbers_2022, xlzdcollaboration2024xlzddesignbooknextgeneration}. While mitigation strategies include purification, event tagging~\cite{aprile2024offlinetaggingradoninducedbackgrounds}, and surface treatments~\cite{copperplating}, the most direct approach remains the screening of construction materials for radon emanation to inform material selection.

Current radon screening strategies can be further strengthened through measurements of radon emanation at cryogenic temperatures, where diffusion is expected to be suppressed in many materials. For reference, radon liquefies at \SI{211.15}{K} and solidifies at \SI{203.15}{K}, while typical liquid xenon (LXe) detectors operate 
at $\sim$\SI{173}{K}.

The Cold Radon Emanation Facility (CREF) at Rutherford Appleton Laboratory (RAL), Oxfordshire, UK, has been developed to perform high-sensitivity radon emanation measurements at temperatures relevant to next-generation rare-event searches. In addition to providing assays under realistic operating conditions, the facility is designed to enable systematic studies of the temperature dependence of radon emanation.
The design builds upon established radon assay systems, such as those developed for the LZ~\cite{Aalbers_2023} and SuperNEMO~\cite{Petro:2025i6} experiments. 


Section 2 provides an overview of the facility, and Section 3 describes the radon detection method employed at CREF. Section 4 presents the first comparative room- and cold-temperature assays of radon emanation from titanium, a material widely used in noble liquid dark matter experiments, including for the LZ cryostat and as a candidate for XLZD.

\section{Facility Overview}
\label{sec:Facility Overview}

\begin{figure}[htbp]
\centering
\includegraphics[width=.7\textwidth]{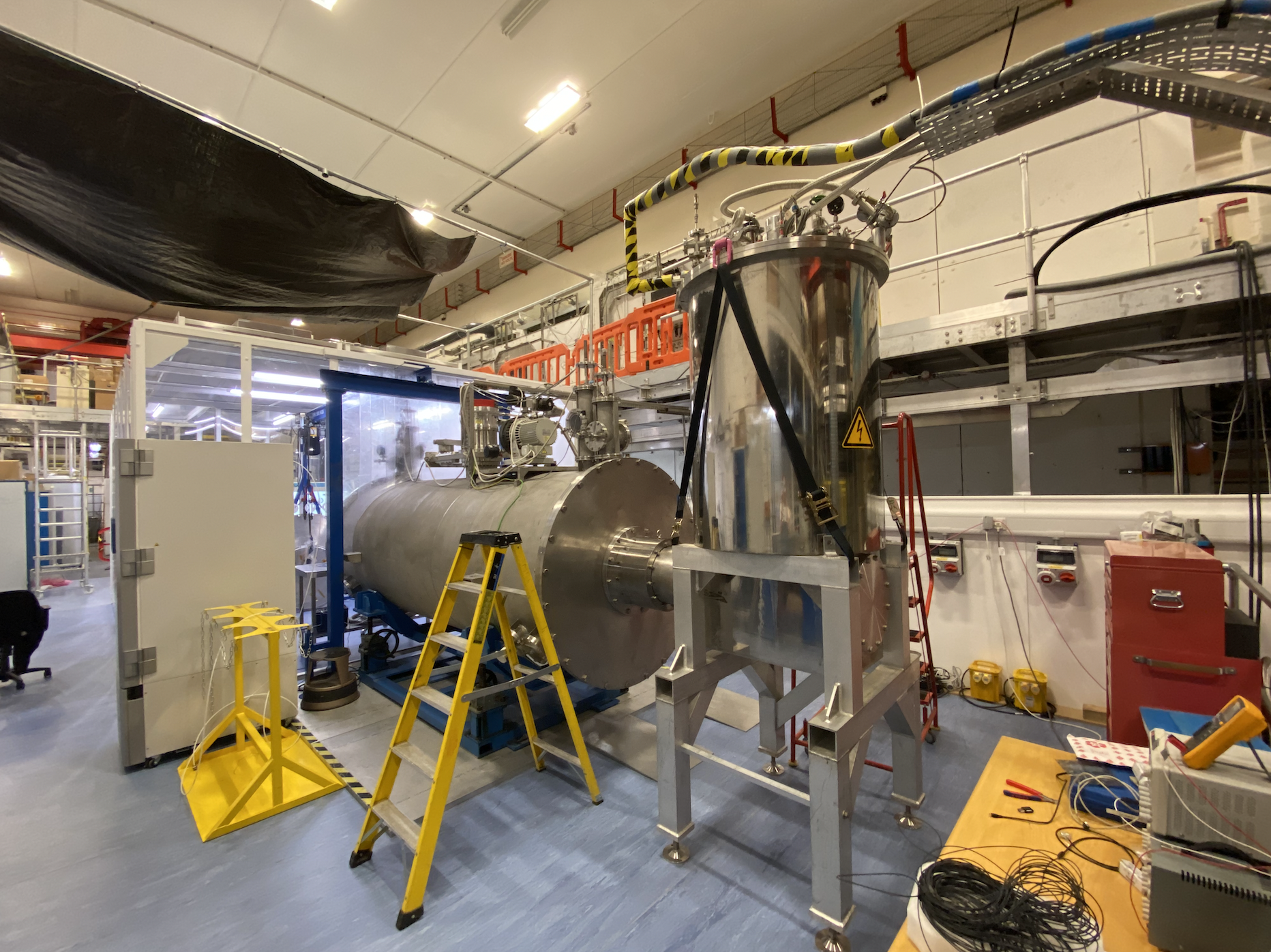}
\qquad
\includegraphics[width=.93\textwidth]{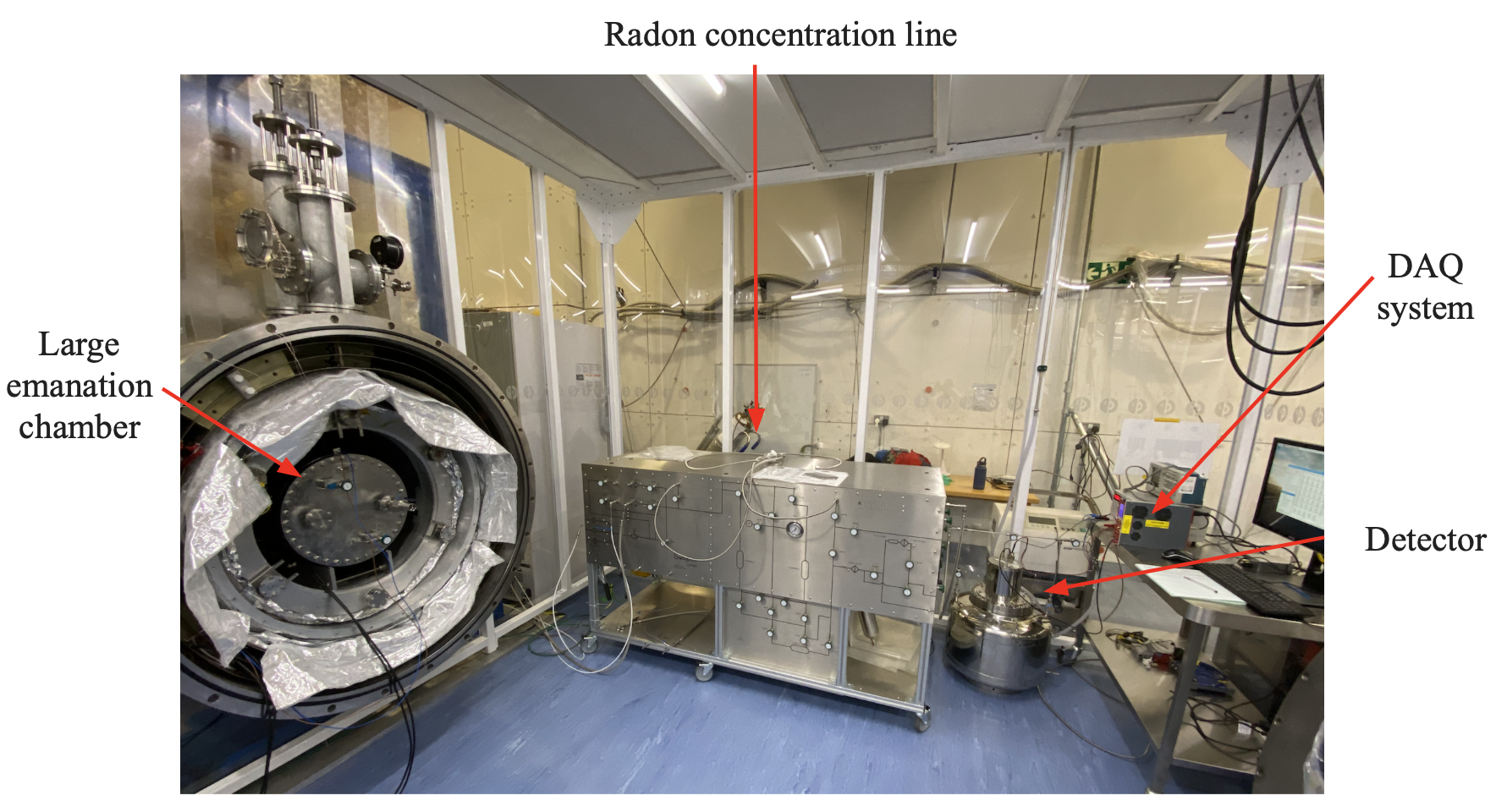}
\caption{Views of the CREF. Top: A photograph of the CREF setup from outside the ISO-6 cleanroom (classified for $<1,000,000$ particles per cubic meter $\ge$\SI{0.1}{\um}~\cite{iso6_ref}). The large emanation chamber, housed within an \SI{800}{L} cryogenic vessel, is shown connected to the cleanroom, with the cooling cryostat on the right. Bottom: An interior photograph of the CREF within the ISO-6 cleanroom. The flange of the large cryogenic vessel has been removed, revealing the large emanation chamber on the left. The visible gas system is employed for the collection, concentration, and transfer of radon samples to the detector, positioned centrally in the image. The detector's DAQ system and associated high-voltage/low-voltage supplies are located on a bench on the right.}
\label{fig:facility_overview}
\end{figure}

The Cold Radon Emanation Facility (CREF) is a dedicated cryogenic radon emanation system comprising two emanation chambers with associated cooling systems (Section~\ref{sec:Emanation Chambers}), a specially developed gas handling system referred to as the radon concentration line (Section~\ref{Radon Concentration Line}) and an electrostatic detector and DAQ system used to measure the radon emanation of a given sample (Section~\ref{Electrostatic Detector Overview}). 
In a typical measurement, samples are sealed within an emanation chamber to allow radon to accumulate, after which the gas is transferred, either directly or via the radon concentration line for larger volumes, to the detector, where the activity is measured via alpha spectroscopy.
A pictorial overview of the facility is given in Figure \ref{fig:facility_overview}. 

\subsection{Emanation Chambers}
\label{sec:Emanation Chambers}

\subsubsection{Small Emanation Chamber}
\label{sec:Small Emanation Chamber}

The facility operates with two emanation chambers. The first is a small cylindrical chamber of \SI{2.7}{L} (\SI{152.4}{mm} length, \SI{146}{mm} diameter), consisting of a stainless steel tubular body sealed by flanges with copper gaskets. It is designed for high-sensitivity radon emanation measurements of small samples, where reduced surface area minimizes intrinsic background. 

The small emanation chamber (SEC) can be operated at room temperature or cooled for measurements at fixed temperatures. Cooling is achieved by direct immersion in an ethanol bath, maintained by an EK90 chiller (down to \SI{183.15}{K}). Thermocouples within the bath monitor the chamber temperature throughout each assay.

\subsubsection{Large Emanation Chamber}
\label{sec:Large Emanation Chamber}

The larger emanation chamber (LEC) is a cylindrical vessel with a volume of \SI{200}{L}, measuring \SI{2.2}{m} in length and \SI{0.35}{m} in diameter. Originally part of the CryoEDM experiment, which sought to measure the neutron's electron dipole moment with a precision of 10$^{-28}e\cdot cm$ \cite*{Cyro_EDM}, this system was acquired to house large samples and construction components for future dark matter experiments, allowing for the determination of their radon emanation rates as a function of temperature. To minimize intrinsic radon activity, its inner surface has been electro-polished for improved material cleanliness, and the vessel itself is constructed from specially selected low-radioactivity stainless steel 304 (sourced from Nironit~\cite{Nironit_ref}) and sealed by a DN350 CF flange. The LEC operates within an \SI{800}{L} cryogenic vessel, capable of reaching liquid nitrogen (LN$_2$) temperatures, $\sim$~\qty{77}{\kelvin}.

\begin{figure}[htbp]
\centering
\includegraphics[width=.4\textwidth]{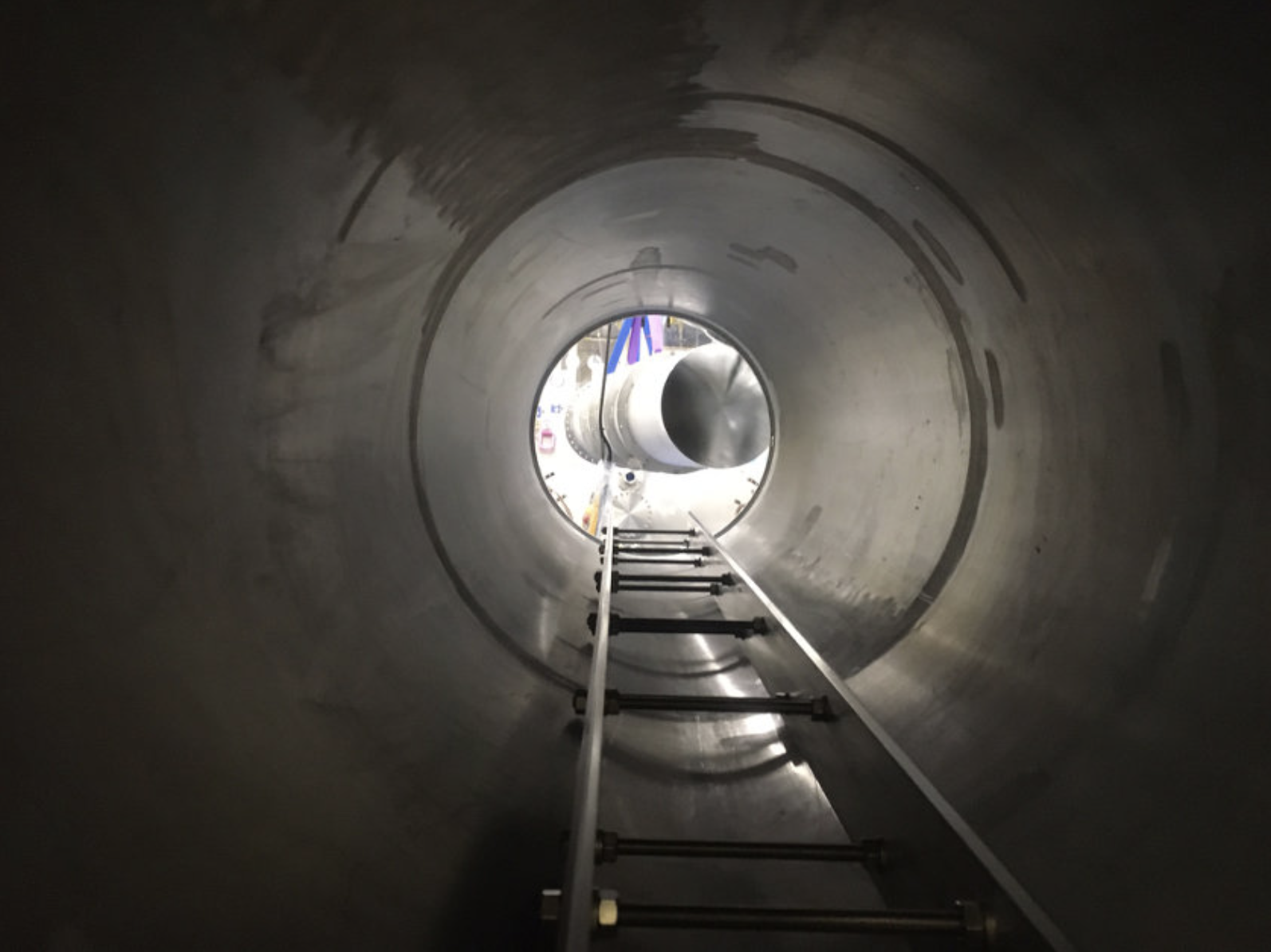}
\qquad
\includegraphics[width=.4\textwidth]{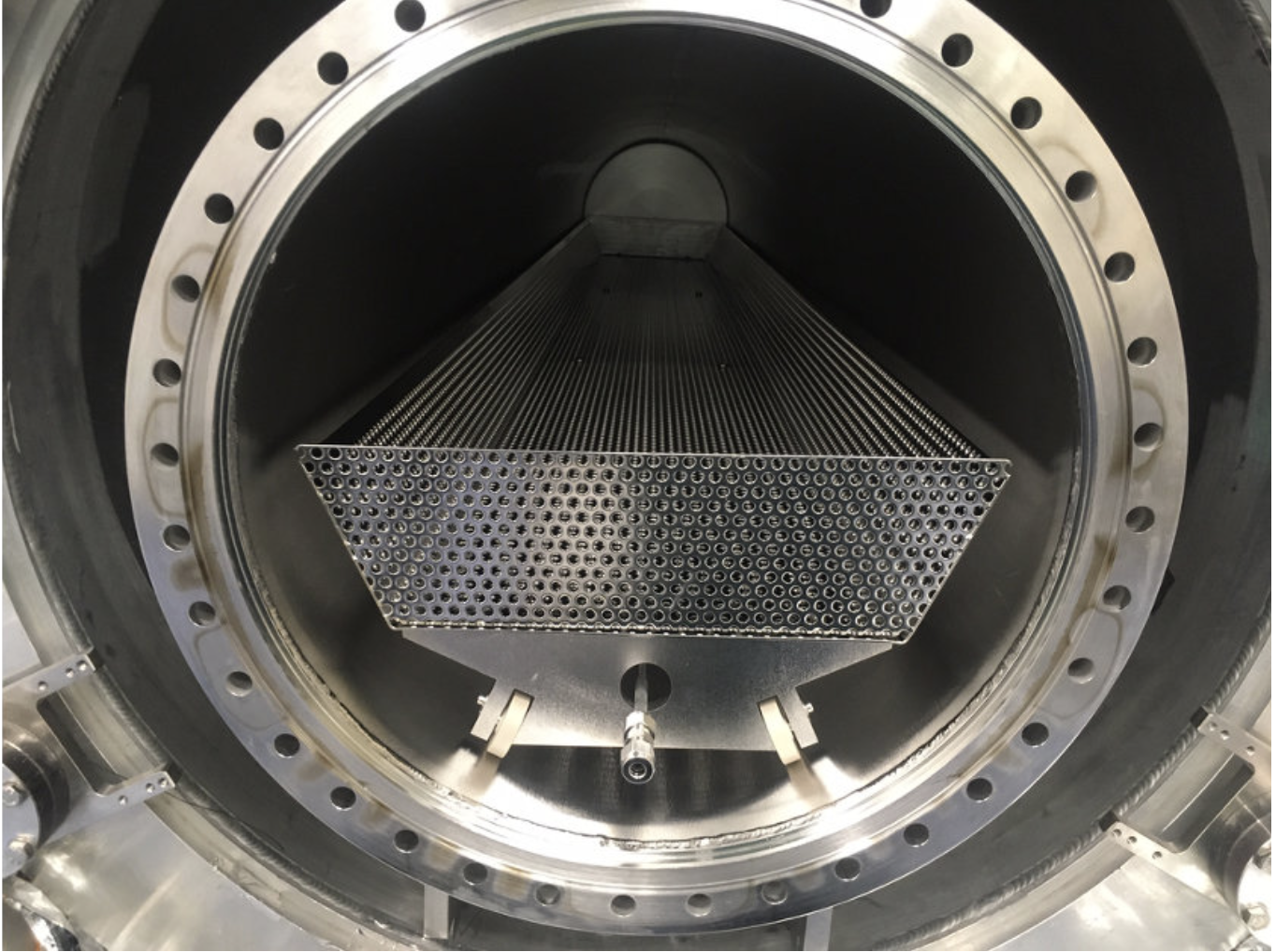}
\caption{The LEC and its cryogenic vessel. Left: Interior view of the \SI{800}{L} cryogenic vessel, captured prior to the installation of the large emanation chamber. Right: Interior of the large emanation chamber, showing the sample tray fully in place with the front gasket and flange removed.
\label{fig:LEC}}
\end{figure}

Both the emanation chamber and the cryogenic vessel are fully assembled and commissioned, as shown in Figures~\ref{fig:facility_overview} and \ref{fig:LEC}. The \SI{800}{L} cryogenic housing comprises inner and outer thermal shields, which are separate from the emanation chamber. Gaseous nitrogen from dewars is circulated through the outer shield and progressively cools the system, allowing liquid nitrogen to accumulate in the inner shield.
The large chamber is mounted on four movable trucks running on rails, enabling near-complete extraction from the cryogenic housing and providing full access to the sealing flange within the ISO-6 cleanroom.

Samples are placed on a removable tray within the emanation chamber, allowing straightforward loading and unloading, as shown in Figure~\ref{fig:LEC}. During measurements, clean carrier gas is flushed through the chamber to extract emanated radon, using a volume approximately ten times that of the chamber. Owing to this large volume, the gas is first passed through the radon trap and subsequently transferred as a concentrated sample to the detector (Section~\ref{Radon Concentration Line}). The determination of the detection efficiency is described in Section~\ref{Detection-efficiency}.

The sensitivity of the large emanation chamber is determined by several factors, including the intrinsic chamber background, the efficiency of radon release from cooled surfaces, and the subsequent concentration and transfer of radon to the detector.
However, studies of the temperature dependence of radon emanation do not necessarily require sub-mBq sensitivity, as higher-activity materials can provide sufficient signal-to-noise.
The facility is therefore designed to operate across a range of measurement regimes, enabling both high-sensitivity assays of low-activity samples and temperature-dependent studies using higher-activity materials, as well as the direct characterization of larger components relevant for next-generation experiments.

\subsection{Radon Concentration Line}
\label{Radon Concentration Line}

\begin{figure}[htbp]
\centering
\includegraphics[width=.8\textwidth]{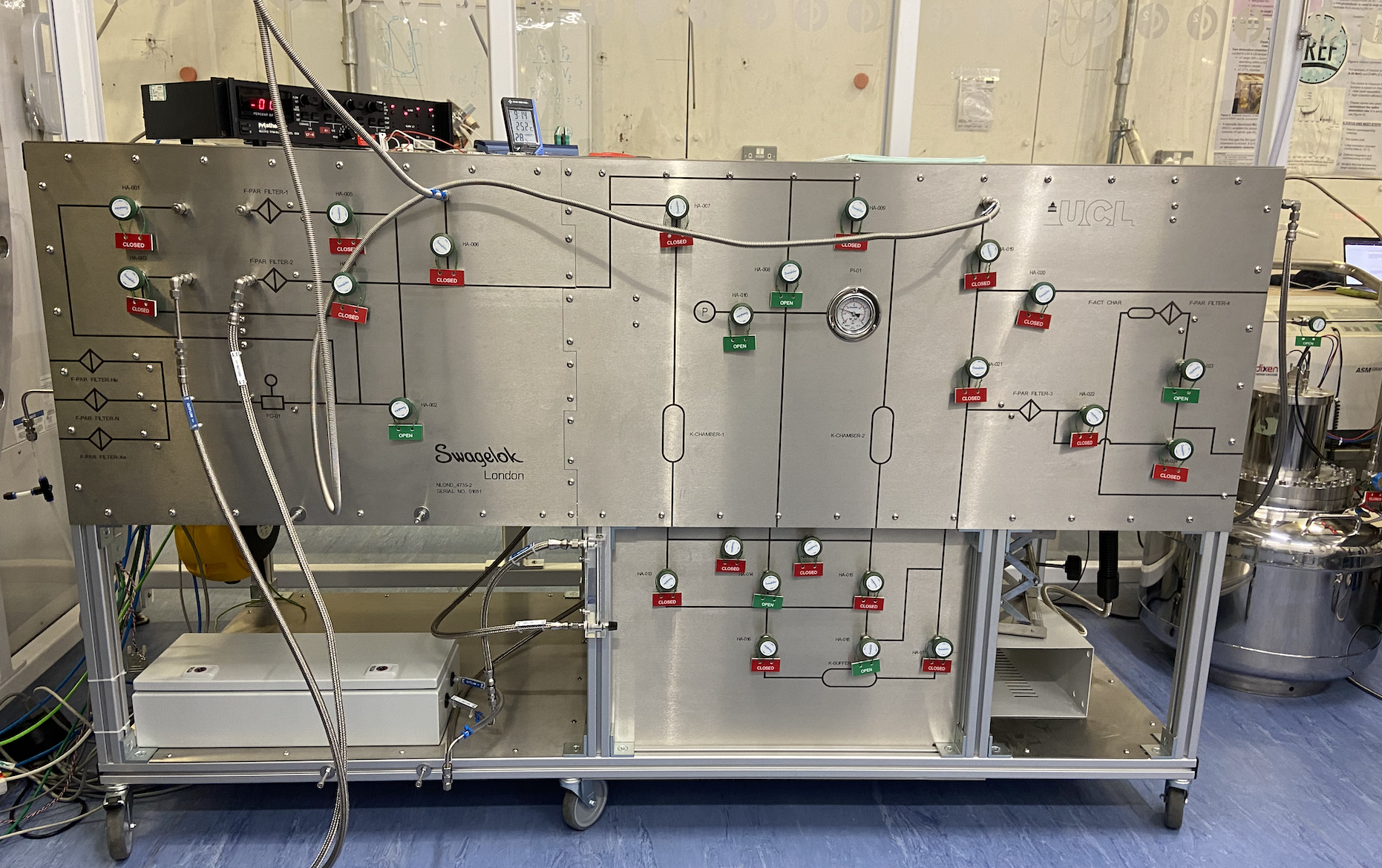}
\qquad
\includegraphics[width=.8\textwidth]{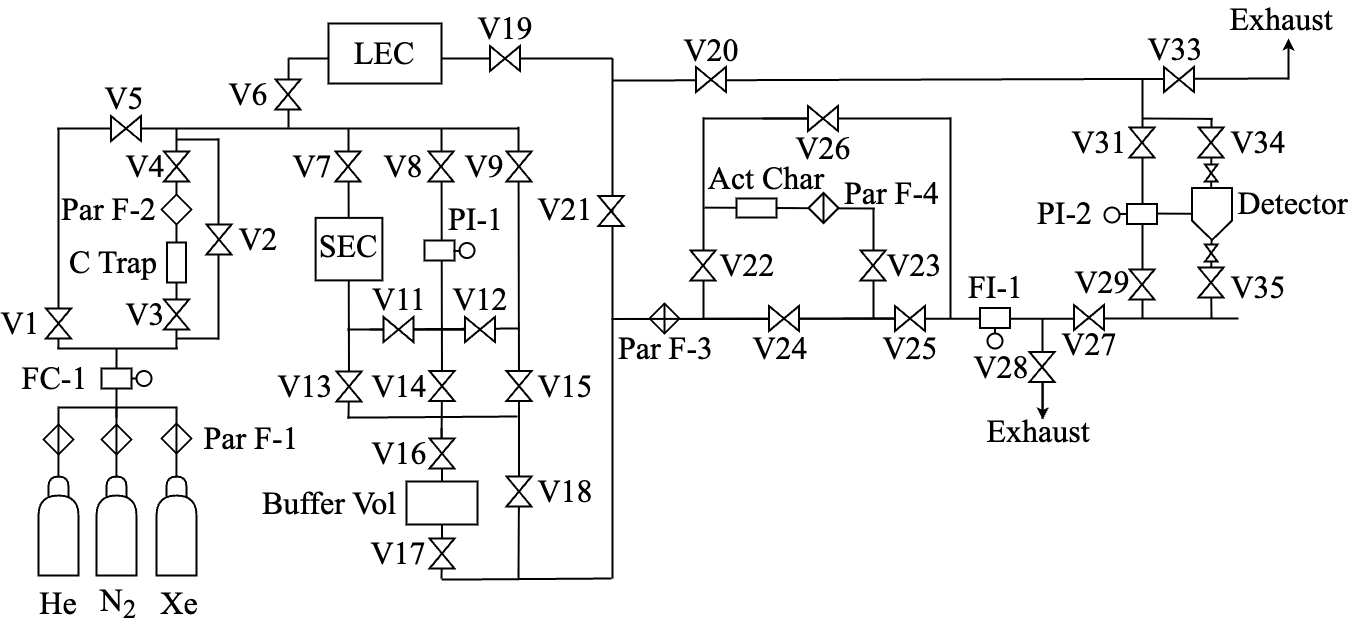}
\caption{An overview of the CREF RnCL. Top: A photograph of the RnCL situated within the cleanroom. Key components visible include the main gas system control panel and the detector located on the right. Bottom: A P$\&$ID illustrating the entire CREF gas system, encompassing the RnCL. The LEC and the SEC are depicted. Additionally, two parallel carbon scrubber cylinders inside the freezer, labeled "C Trap", responsible for removing impurities from the carrier gas supplied by storage bottles are shown. The system can be operated with helium, xenon and nitrogen gases, with the latter being the default. The "Act Char" activated charcoal trapping unit, utilized during LEC operation, is also included in this diagram.
 \label{fig:RnCL_diagram}}
\end{figure}

A radon harboring system known as a radon concentration line (RnCL), illustrated in Figure~\ref{fig:RnCL_diagram}, was developed at University College London (UCL) to characterize backgrounds for the SuperNEMO neutrinoless double beta-decay experiment~\cite{Mott2013}. A modified version of the RnCL is employed at CREF to enable high-sensitivity radon measurements from large gas volumes, through concentration and transfer of radon to the detector in conjunction with the emanation chambers.

The RnCL is designed for flexible operation within CREF, supporting different carrier gases, chamber configurations, and radon transfer schemes. In this work, measurements were performed using the small emanation chamber (SEC), for which radon samples were transferred directly to the detector without use of the RnCL.

For larger volumes, such as those associated with the large emanation chamber (LEC), a two-step transfer is required. In this mode, radon is carried by a gas flow through the RnCL and trapped on an activated charcoal unit ($\sim$\SI{200}{g} CarboAct labeled as "Act Char" in Figure~\ref{fig:RnCL_diagram}). The trap is maintained at \SI{245}{K} to enhance adsorption, and subsequently heated to \SI{493}{K} to release a concentrated radon sample for transfer to the detector.

The overall performance of the system depends on the efficiency of radon transfer and detection. In this work, the detector response is calibrated as described in Section~\ref{Detection-efficiency}.

To minimize backgrounds, the RnCL is constructed from low-radioactivity materials, with stainless steel components throughout and zirconiated welding rods used in place of thoriated alternatives. Carrier gas is purified prior to use by particle filtration and passage through a charcoal-based radon scrubber operated at \SI{193}{K} (labeled "C-Trap" in Figure~\ref{fig:RnCL_diagram}).

\section{Radon Detection Method}
\label{sec:Radon Detection Method}

\subsection{Methodology}
\label{Methodology}

To determine the radon emanation rate, samples are placed in the small emanation chamber (SEC, \SI{2.7}{L}), which is filled with nitrogen carrier gas at atmospheric pressure and sealed to allow radon to accumulate over a defined period. Following accumulation, the gas sample is transferred to an electrostatic detector, where the decay of radon progeny is measured via alpha spectroscopy.

The detector is sensitive to the alpha decays of $^{218}$Po and $^{214}$Po, which are produced in the $^{222}$Rn decay chain. The measured count rates of these isotopes are related to the underlying radon activity through the time evolution of the decay chain. Rather than directly measuring $^{222}$Rn, the radon activity is therefore inferred from the observed polonium populations.

\begin{figure}[htbp]
\centering
\includegraphics[width=.7\textwidth]{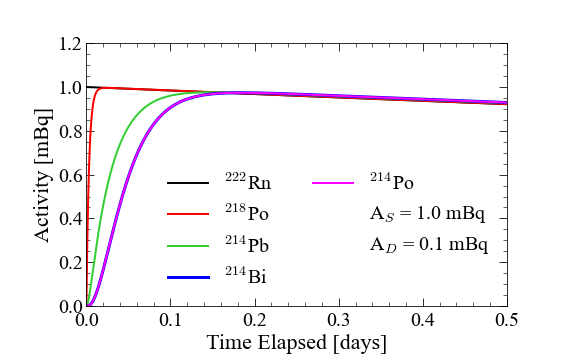}
\caption{Time evolution of the activities of isotopes in the $^{222}$Rn decay chain following the introduction of $A_S = 1.0$~mBq of $^{222}$Rn into a detector with a constant background activity of $A_D = 0.1$~mBq. 
 \label{fig:activity_example}}
\end{figure}

The time evolution of the number of atoms $N_i$ of each isotope in the chain is governed by

\begin{equation}
    {\frac{dN_i}{dt}} = {\lambda_{i-1}N_{i-1} - \lambda_{i}N_{i}}.
    \label{rn_decay_1}
    \end{equation}

where $\lambda_i$ is the decay constant of isotope $i$. The resulting activity evolution of the radon progeny following injection into the detector is described by the standard Bateman equations~\cite{Mott2013}. An example of the expected time dependence of the relevant isotopes is shown in Figure~\ref{fig:activity_example}, illustrating the rapid equilibration of $^{218}$Po with $^{222}$Rn and the delayed growth of $^{214}$Po.

In this work, the measured time-dependent rates of $^{218}$Po and $^{214}$Po are fitted using these decay-chain relationships to extract the initial radon activity in the detector. The reconstructed activity is corrected for detector and transfer efficiencies, which are determined independently using calibrated $^{226}$Ra sources, as described in Section~\ref{Detection-efficiency}.

\subsection{Electrostatic Detector Overview}
\label{Electrostatic Detector Overview}

\begin{figure}[htbp]
\centering
\includegraphics[width=.4\textwidth]{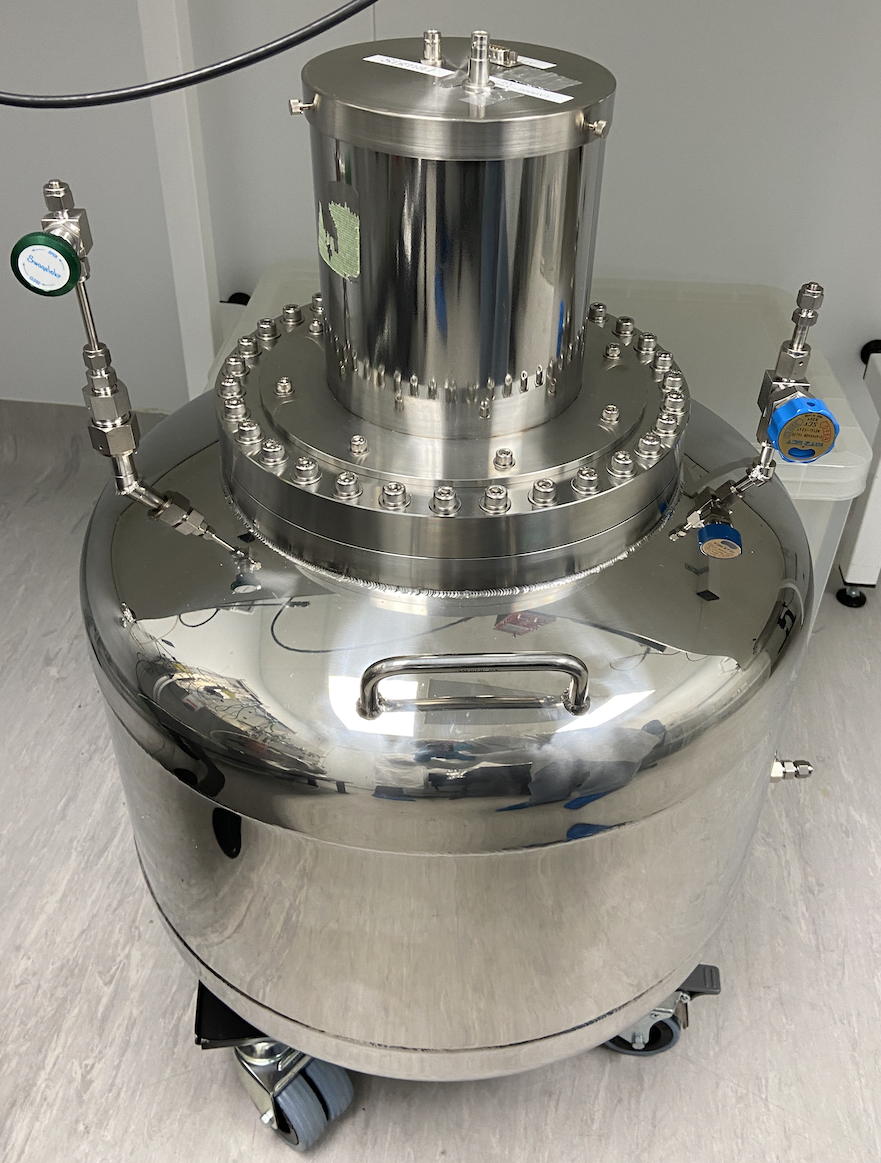}
\qquad
\includegraphics[width=.4\textwidth]{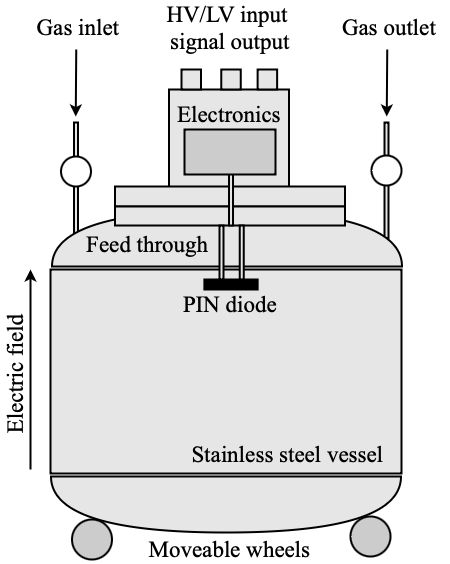}
\caption{Left: Photograph of the CREF electrostatic detector. Right: Schematic overview of the detector, highlighting the gas inlet and outlet, the electronics housed within a Faraday cage at the top, and the PIN diode located inside the stainless steel vessel and connected via a feed-through system.
}
\label{fig:detector}
\end{figure}

The radon activity is measured using an electrostatic detector developed by the University of Tokyo and supplied by CosmoTec~\cite{Cosmotec_ref}. This class of detector has been widely used for low-level radon measurements in rare-event experiments, including earlier implementations for ELEGANT V and Super-Kamiokande, as well as more recent developments~\cite*{Choi_2001, Mitsuda_2003, Pronost_2018}. The detector consists of a \SI{70}{L} stainless steel vessel equipped with a silicon PIN diode as shown in Figure~\ref{fig:detector}.


The detector is operated with an applied high voltage of \SI{-1900}{V}. This voltage is divided between the PIN diode bias and the electrostatic collection field, with \SI{-100}{V} applied to reverse-bias the PIN diode and \SI{-1800}{V} used to establish the electric field within the detector volume.

Following introduction of the sample gas, positively charged radon progeny are drifted onto the diode surface. A large fraction of these progeny are produced as positive ions (measured to be $(87.3 \pm 1.6)\%$ for $^{218}$Po in air~\cite*{po_rates_nitrogen}), enabling efficient electrostatic collection. The subsequent alpha decays are detected via their energy deposition in the diode.

The alpha peaks corresponding to $^{218}$Po and $^{214}$Po are well separated in energy, allowing their activities to be independently measured. In this work, the $^{214}$Po channel is used for activity reconstruction due to its higher stability and reduced sensitivity to neutralization effects.

The detector is operated with a charge-sensitive preamplifier and digitized using a CAEN waveform digitizer. Pulse processing is performed using a digital trapezoidal filter to extract alpha energies. The resulting time-dependent count rates are used to reconstruct the radon activity as described in Section~\ref{Methodology}.

\subsection{Detection Efficiency}
\label{Detection-efficiency}

To determine the detection efficiency of the electrostatic detector, a calibrated radon source is used in a dedicated measurement. A $^{226}$Ra flow-through source (Pylon Electronics, RN-1025) is employed to introduce a known quantity of $^{222}$Rn into the detector volume.

Prior to measurement, the source is flushed with nitrogen to remove residual activity. It is then sealed and allowed to accumulate radon over a time $T$. The activity of $^{222}$Rn generated during this period is given by



\begin{equation}
\label{eq:activityequation}
{A\left(^{222}\text{Rn}\right)} = {A\left(^{226}\text{Ra}\right)}\left(1-e^{-T\lambda}\right),
\end{equation}

where $A(^{226}\mathrm{Ra})$ is the source activity and $\lambda$ is the decay constant of $^{222}$Rn. A correction is applied to account for the decay occurring during transfer of the radon to the detector.

\begin{figure}[htbp]
\centering
\includegraphics[width=.7\textwidth]{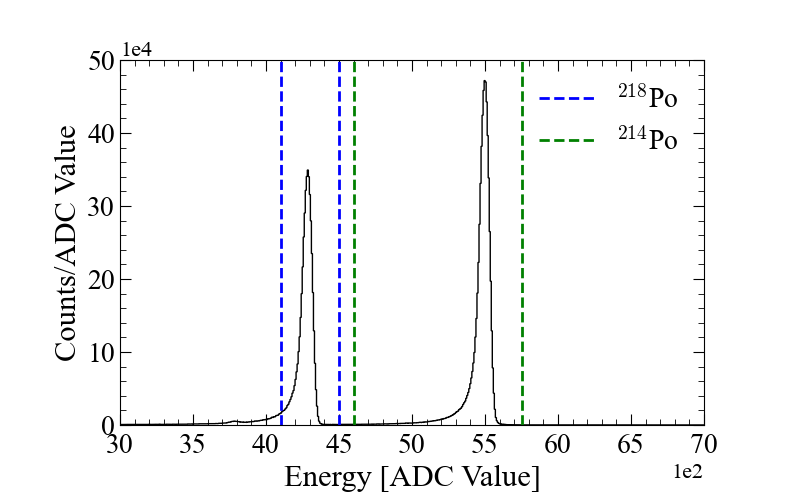}
\caption{
Alpha energy spectrum from a CREF detector efficiency calibration. The alpha energy spectrum was obtained after introducing \SI{12.62}{Bq} of $^{222}$Rn into the detector. Indicated peaks represent $^{218}$Po (blue) and $^{214}$Po (green). The respective energy cuts, expressed in ADC channels, used to isolate the  peaks are 4100-4500 [ADC Value] for $^{218}$Po and 4600-5750 [ADC Value] for $^{214}$Po.
\label{fig:h1_spectrum_run_22}}
\end{figure}

\begin{figure}[htbp]
\centering
\includegraphics[width=.7\textwidth]{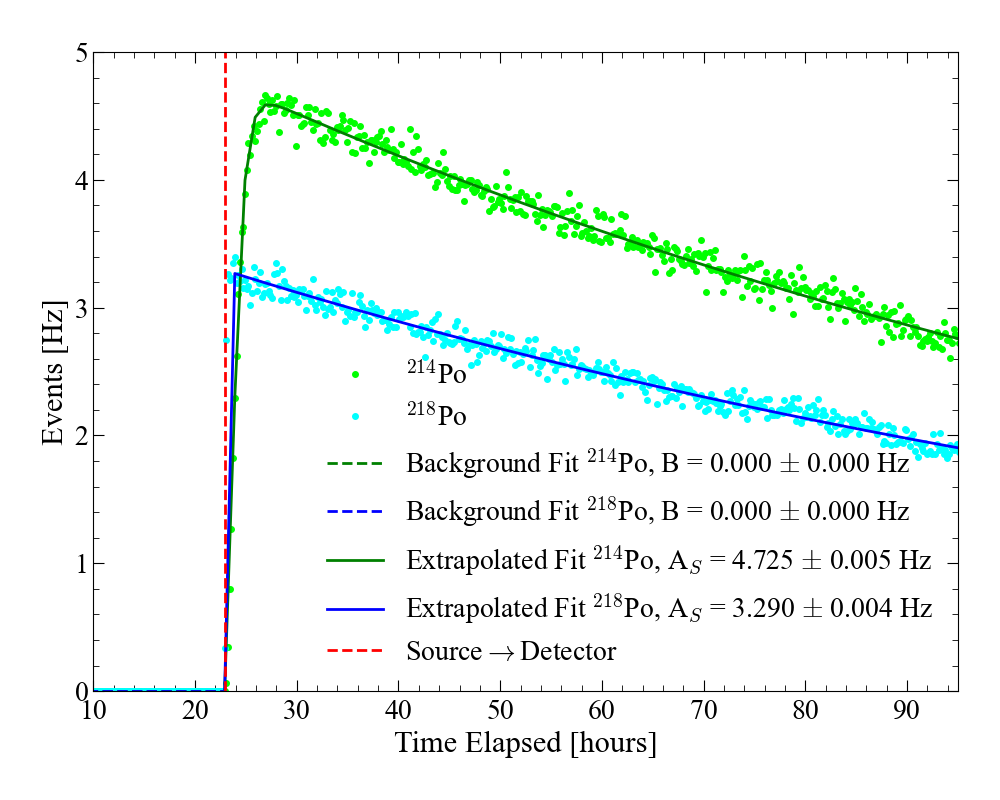}
\caption{Detected event rate of $^{214}$Po (green points) and $^{218}$Po (blue points) following the introduction of \SI{12.62}{Bq} of $^{222}$Rn into the detector (red dashed line). The data are fitted with iterative decay models (solid lines), which are extrapolated to determine the source activity $A_S$. Background levels are verified to be negligible prior to injection using flat fits in the $^{218}$Po (blue dashed line) and $^{214}$Po (green dashed line) regions. Statistical uncertainties are not shown.
\label{fig:Po214andPo218SpikePlot_run_22}}
\end{figure}

Following injection, the time-dependent rates of $^{218}$Po and $^{214}$Po are measured. The corresponding alpha energy spectrum is shown in Figure~\ref{fig:h1_spectrum_run_22}, where the isotopes are identified through distinct energy windows. The temporal evolution of their activities is shown in Figure~\ref{fig:Po214andPo218SpikePlot_run_22} and is fitted using the decay-chain model described in Section~\ref{Methodology}, enabling extraction of the initial radon activity in the detector.


The detection efficiency $\epsilon$ is defined as the ratio of the measured activity to the known injected activity. In this work, efficiencies of $(26.1 \pm 1.0)\%$ for $^{218}$Po and $(37.4 \pm 1.5)\%$ for $^{214}$Po are obtained. The dominant uncertainty arises from the quoted activity of the calibration source.

The lower efficiency observed for $^{218}$Po is consistent with its known susceptibility to neutralization by gas impurities, which reduces the fraction of ions available for electrostatic collection. In contrast, $^{214}$Po provides a more stable and reliable measure of the radon activity and is therefore used for all subsequent analyses.

The maximum achievable efficiency for this class of detector is limited to approximately 50\% due to geometric considerations, as only a fraction of emitted alpha particles are directed towards the diode.

\subsection{Minimum Detectable Activity}
\label{Minimum Detectable Activity}

The sensitivity of the detector is quantified through the minimum detectable activity (MDA), defined as the smallest activity distinguishable from background at a given confidence level.

Following injection, no further radon is introduced into the detector and the number of radon atoms evolves according to



\begin{equation}
\label{eq:activityequation2}
\frac{dN}{dt} = - \lambda N,
\end{equation}

where $\lambda$ is the decay constant of $^{222}$Rn. For an initial activity $A_S$ at $t=0$, the time-dependent activity is therefore


\begin{equation}
\label{eq:activityequation3}
A(t) = A_S e^{-\lambda t}.
\end{equation}

The measured count rate is given by the sum of the decaying signal and a constant background contribution,


\begin{equation}
\label{eq:countrate}
R(t) = \epsilon A_S e^{-\lambda t} + \epsilon A_D,
\end{equation}

where $\epsilon$ is the detection efficiency and $A_D$ is the background activity determined from dedicated measurements.

The expected number of signal events over a measurement time $T$ is


\begin{equation}
\label{eq:signal}
S = \epsilon \int_0^T A_S e^{-\lambda t}  dt = \epsilon \frac{A_S}{\lambda}\left(1 - e^{-\lambda T}\right),
\end{equation}

while the expected number of background events is

\begin{equation}
\label{eq:bkg}
B = \epsilon A_D T.
\end{equation}

The minimum detectable signal $S_0$ at a given confidence level is determined using a normal approximation to Poisson counting statistics~\cite*{G_F_Knoll},

\begin{equation}
\label{eq:min-signal}
S_0 = 2E\left(E + \sqrt{2B}\right),
\end{equation}

where $E = \mathrm{erf}^{-1}(2\mathrm{CL} - 1)$. The MDA corresponds to the value of $A_S$ for which $S = S_0$.

\begin{figure}[htbp]
\centering
\includegraphics[width=.7\textwidth]{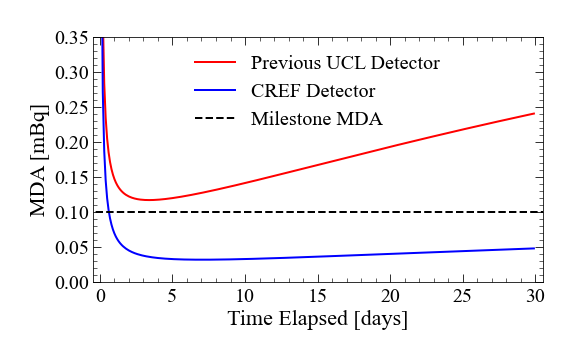}
\caption{Minimum detectable activity (MDA) of the CREF detector as a function of measurement time (blue line), shown at 90\% confidence level. The performance of the previous UCL radon detector used for LZ and SuperNEMO is also shown (red line)~\cite{ScreeningPaper2020}. The CREF design target of \SI{0.1}{mBq} is indicated by the black dashed line.
\label{fig:CREF_h2_MDA_time}}
\end{figure}

Using the measured detector efficiency  ($\epsilon \approx 37.4\%$ for $^{214}$Po) and background rate  ($\sim 0.286$ cpd in the $^{214}$Po analysis window), the resulting MDA at 90\% CL as a function of measurement time is shown in Figure~\ref{fig:CREF_h2_MDA_time}. The detector achieves a sensitivity at or below \SI{0.05}{mBq} for typical assay durations, improving on the performance of a previous detector iteration, as shown in Figure~\ref{fig:CREF_h2_MDA_time}. 




\section{Temperature-Dependent Radon Emanation Measurements}
\label{sec:Room-Cold Temperature Comparative Sample Assays}

\subsection{SEC Measurements}
\label{SEC Only}

Prior to sample measurements, dedicated background runs were performed using the SEC at both room temperature (\SI{293}{K}) and reduced temperature (\SI{223}{K}). Before counting, the chamber was flushed with clean carrier gas, and sealed for an extended period (30 days) to allow radon to accumulate towards equilibrium conditions. These measurements establish the intrinsic background of the emanation chamber and detector system under the relevant operating conditions.

From these background measurements, the internal emanation of the chamber-detector system, as inferred from the $^{214}$Po channel, was found to be $0.104 \pm 0.633$~cpd at \SI{293}{K} and $3.378 \pm 0.998$~cpd at \SI{223}{K}. The corresponding $^{218}$Po rates were $2.031 \pm 0.704$~cpd and $3.350 \pm 1.044$~cpd, respectively. A difference at the $\sim 2.8\sigma$ level is observed in the $^{214}$Po background rate between room and reduced temperature.

Given the relatively large statistical uncertainties and the consistency of the $^{218}$Po measurements, this difference may arise from statistical fluctuations or residual systematic effects. Background levels were determined independently for each temperature configuration and subtracted on a per-measurement basis, such that this variation does not bias the extracted sample activities.

These results demonstrate that the background of the SEC-detector system remains at the level of a few counts per day across the temperature range considered, establishing a sufficiently low baseline for subsequent radon emanation measurements.

\subsection{Titanium Samples}
\label{Titanium Samples}

Titanium has conventionally been used for cryostat vessels in noble liquid dark matter experiments, where it constitutes a substantial fraction of the detector mass and surface area and is in close proximity to the active volume. As such, dedicated screening campaigns have been undertaken, for example by LZ, to ensure that selected titanium exhibits ultra-low levels of radioactive contamination~\cite{ScreeningPaper2020}. Despite these efforts, LZ observed a discrepancy between the predicted $^{222}$Rn activity and a higher in-situ measured value during its initial science run. One possible explanation for this excess is the presence of contamination at or near the titanium surface, potentially introduced during fabrication, handling, or cleaning processes~\cite{Aalbers_2022}. Additionally, effects such as non-uniform radium distributions or enhanced near-surface diffusion may contribute to increased emanation. These considerations motivate further study of radon emanation in titanium, particularly under detector-relevant conditions.

\begin{figure}[htbp]
\centering
\includegraphics[width=.7\textwidth]{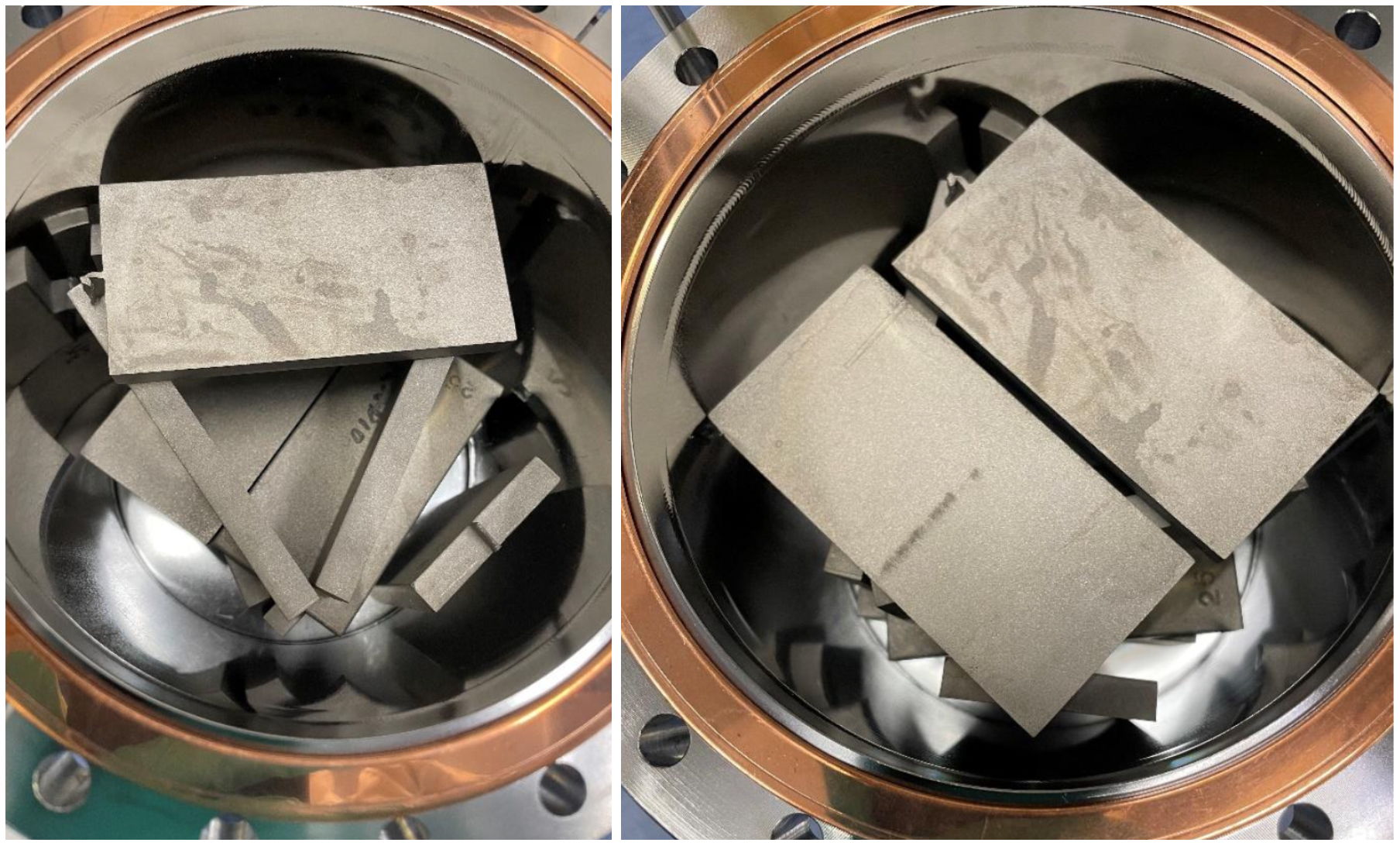}
\caption{Photographs of the titanium samples arranged inside the SEC prior to measurement. Left: Partially assembled configuration before placement of the final sample. Right: Final arrangement of all 11 samples inside the chamber prior to the emanation period.
\label{fig:titanium_combined}}
\end{figure}

For this reason, titanium was selected as the first material for comparative room- and cold-temperature assays at CREF. A total of 11 titanium cut-outs were obtained from Loterios~\cite{Loterios_ref} as part of the LZ screening campaign~\cite{ScreeningPaper2020}. This material was not selected for cryostat construction due to its comparatively high activity. The samples had a combined mass of $2288.4 \pm 0.5$~g and a total surface area of $12700 \pm 317$ mm$^2$. Typical sample dimensions were $100 \times 50 \times 9$ mm$^3$, with an uncertainty of \SI{1}{mm} on each dimension. The samples were cleaned using isopropanol, including ultrasonic agitation, but were not subjected to chemical etching. As shown in Figure~\ref{fig:titanium_combined}, visible surface blemishes and markings remain. The visible surface condition suggests that surface-related contributions may play a significant role in the measured emanation.


Assay measurements were performed at one room temperature (\SI{293}{K}) and two reduced temperatures (\SI{223}{K} and \SI{188}{K}). For each measurement, the SEC was flushed at \SI{293}{K} with more than 10 chamber volumes of carrier gas and subsequently sealed at atmospheric pressure. Radon was allowed to accumulate for approximately 30 days. For cold measurements, the accumulation period commenced only after the target temperature had been reached and stabilized. Temperature monitoring of the ethanol bath indicated fluctuations less than $\pm$~\SI{2}{K} during the measurement period.

\begin{figure}[htbp]
\centering
\includegraphics[width=.7\textwidth]{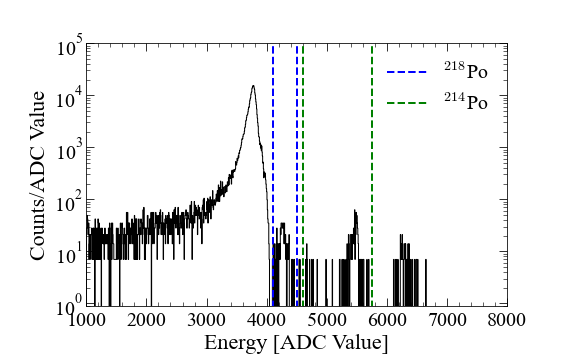}
\caption{Alpha energy spectrum of the titanium sample measured at \SI{293}{K} using the SEC. The energy regions used to isolate the $^{214}$Po (green dashed lines) and $^{218}$Po (blue dashed lines) peaks are indicated. The intrinsic $^{210}$Po background peak is clearly visible at $\sim$3700 ADC channels.
\label{fig:h1_spectrum_run_38_39}}
\end{figure}
\begin{figure}[htbp]
\centering
\includegraphics[width=.61\textwidth]{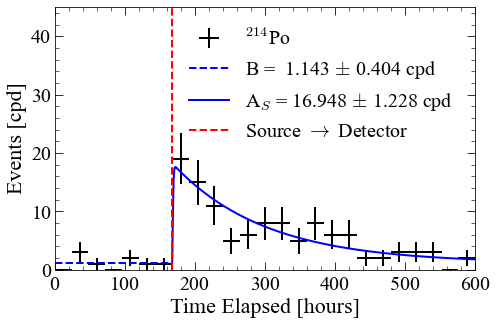}
\caption{Observed $^{214}$Po rate during the \SI{293}{K} titanium sample measurement with the SEC. The time at which the sample is transferred from the SEC to the detector is indicated by the red dashed line.\label{fig:PoSpikePlots_run_38_39}}
\end{figure}

The radon activity of the titanium samples was determined from the $^{214}$Po channel after applying detector efficiency corrections, and the results are summarized in Table~\ref{tab:titanium_assay_results_corrected}. The measured activity at \SI{293}{K} is consistent with previous measurements of similar Loterios titanium samples performed at Boulby~\cite*{ScreeningPaper2020}. The corresponding alpha energy spectrum and time evolution of the $^{214}$Po rate for the \SI{293}{K} measurement are shown in Figures~\ref{fig:h1_spectrum_run_38_39} and~\ref{fig:PoSpikePlots_run_38_39}, respectively.


A clear suppression of radon emanation is observed at reduced temperature. At \SI{188}{K}, which is close to the operating temperature of liquid xenon detectors (\SI{173}{K}), a reduction of $52.1 \pm 4.9\%$ relative to the \SI{293}{K} measurement is measured using the $^{214}$Po activity. The suppression factors for both reduced temperatures are summarized in Table~\ref{tab:titanium_assay_results_cold_supression_factors}. Measurements based on the $^{218}$Po channel are not included due to potential contamination from the nearby $^{210}$Po peak (Figure~\ref{fig:h1_spectrum_run_38_39}), and the reduced collection efficiency associated with neutralization effects, as discussed in Section~\ref{Detection-efficiency}. 

This constitutes the first demonstration of a measurement of the cold suppression of radon emanation in titanium, a key material in dark matter and other rare-event search experiments. The specific activity is likely dominated by surface-related contributions rather than intrinsic bulk emanation. 
Such contributions enhance the sensitivity to temperature-dependent effects and thereby facilitate the observation of the suppression.
Future measurements combining radon emanation with gamma spectroscopy, and comparing surface-treated and untreated samples, will be required to disentangle these effects.


\begin{table}[htbp]
\centering
\caption{Results of the SEC titanium sample assays at room and reduced temperatures, with detector efficiency corrections applied.\label{tab:titanium_assay_results_corrected}}
\smallskip
\begin{tabular}{c|c}
\hline
Temperature [K] & $^{214}$Po [mBq/kg] \\
\hline
293 & 41.14 $\pm$ 2.16 \\
223 & 20.20 $\pm$ 2.17 \\
188 & 19.72 $\pm$ 1.53 \\
\hline
\end{tabular}
\end{table}

\begin{table}[htbp]
\centering
\caption{Summary of the suppression in $^{214}$Po activity for the titanium samples at reduced temperatures, expressed as a percentage relative to the 293~K measurement. \label{tab:titanium_assay_results_cold_supression_factors}}
\smallskip
\begin{tabular}{c|c}
\hline
Temperature [K] & $^{214}$Po Suppression [$\%$] \\
\hline
223 & 50.90 $\pm$ 6.09 \\
188 & 52.1 $\pm$ 4.9 \\
\hline
\end{tabular}
\end{table}

\section{Conclusions and Future Outlook} 
\label{sec:Conclusions}

The commissioning and initial operation of the Cold Radon Emanation Facility (CREF) have been presented. The facility has demonstrated the capability to perform high-sensitivity radon emanation measurements at both room and cryogenic temperatures, with the electrostatic detector reaching a minimum detectable activity of $\sim$\SI{0.05}{mBq} at 90\%~CL for typical assay durations.

Comparative measurements of titanium samples at \SI{293}{K} and reduced temperatures (\SI{223}{K} and \SI{188}{K}) show a clear suppression of radon emanation by approximately a factor of two at cryogenic temperatures. This constitutes the first direct observation of temperature-dependent suppression of radon emanation in titanium, a key material for rare-event search experiments. The specific activity observed is likely influenced by surface-related contributions, which in this context enhance the sensitivity to temperature-dependent effects.

These results establish the capability of CREF to perform temperature-dependent radon emanation studies under detector-relevant conditions. Future measurements will aim to disentangle bulk and surface contributions by combining emanation assays with complementary techniques such as gamma spectroscopy, and by comparing samples with controlled surface treatments. Additional studies will extend the temperature range and investigate radon transport and trapping effects within the system.

\bibliographystyle{JHEP}
\bibliography{biblio.bib}








\end{document}